\documentclass[preprint,aps,showpacs,nofootinbib,preprintnumbers,amsmath,amssymb]{revtex4-1}
\usepackage{epsfig}
\usepackage{bm}
\usepackage[usenames ,dvipsnames]{xcolor}
\usepackage{slashed}
\usepackage{graphicx,color}

\usepackage{multirow}
\usepackage{tabularx}

\begin{document}
	
	\title{CP asymmetries  in $B$ meson two-body baryonic  decays}

\author{Chao-Qiang Geng$^1$, Xiang-Nan Jin$^{1,2,3}$ and  Chia-Wei Liu$^{1,3}$ }
\affiliation{
$^1$School of Fundamental Physics and Mathematical Sciences, Hangzhou Institute for Advanced Study, UCAS, Hangzhou 310024, China\\
$^2$University of Chinese Academy of Sciences, Beijing 100049, China \\
$^3$Institute of Theoretical Physics, Chinese Academy of Sciences, Beijing 100190, China \\
$^4$Tsung-Dao Lee Institute \& School of Physics and Astronomy, \\
Shanghai Jiao Tong University, Shanghai 200240, China
}\date{\today}

\begin{abstract}

We  study the CP-odd and CP-even observables of the $B$ mesons decaying into a baryon and antibaryon.
We estimate these observables through the $^3P_0$ model and chiral selection rule. The decay branching ratios of $ B^+ \to p \overline{\Lambda}$ and $ B^0 \to p \overline{p}$ are calculated to be  $2.31 \times 10^{-7}$ and  $1.27 \times 10 ^ { -8} $, which are consistent with the current experiments, respectively. 
The effects of 
the $B-\overline{B}$ oscillations are considered, which largely suppress the direct CP asymmetries in the $B_s^0$ decays. 
We suggest the experiments to visit $B_s^0 \to 
\Lambda(\to p \pi^-)
\overline{\Lambda} (\to \overline{ p}  \pi^+) $, where the time-averaged CP-odd observables are estimated to be  large. The direct CP asymmetries of  $B^+ \to p \overline{\Lambda}$ and $B^0 \to p\overline{p}$ are found to be 
$26.2\%$ and $-3.1\%$ for a positive strong phase and $-36.9\%$ and $4.2\%$ for a negative strong phase, respectively. 
\end{abstract}

\maketitle

\section{Introduction}
CP asymmetries play an important role in the study of particle physics~\cite{CPNP}. 
In particular, 
CP violation  is not only  the key to understand the matter-antimatter asymmetry in the universe but an important way to probe the effects of new physics~(NP). 
As the CP symmetry is respected by the strong interaction, the study of CP violation allows us to extract the complex phases  in the weak interaction of the standard model~(SM)~\cite{Gronau:1990ra,CKM,LHCb:2017hkl,Geng:2022osc} even though theoretical calculations are clouded by the hadronic uncertainties.

Since CP violation was first observed in 2001~\cite{BCP}, various CP-odd quantities have  been measured  mainly in the $B$ meson decays~\cite{pdg}. Remarkably, the LHCb collaboration has recently found evidences of CP violation in $D$ meson decays~\cite{LHCb:2014kcb}, which are expected to be small. The finding has stimulated great interests among the theorist and a debate on whether the observed values require NP is still ongoing~\cite{DmesonCP}.

Despite the great processes in the mesonic sector, a nonzero signal at $5\sigma$ confidential level  of CP violation is still absent in the baryonic sector yet.  In particular, the direct CP violation of  $\Lambda_b \to \Lambda K^- \pi^+$ is found to be $-0.53 \pm 0.25$~\cite{LHCb:2016rja}, which has a huge central value but a large uncertainty. In addition,
several theoretical works~\cite{PQCD} have been triggered by the measurements of $A_{CP}( \Lambda_b \to p \pi^- /K^-)$, which are consistent with 
the results in the perturbative QCD (pQCD)~\cite{private}. 
Undoubtedly, more experiments are expected in the future. 

To the end of probing CP violation involving baryons, we study   the CP asymmetries exhibit in $B\to {\bf B}_1 \overline{{\bf B}}_2$ with ${\bf B}_1$ and $\overline{{\bf B}}_2$  baryon and antibaryon, respectively. On the one hand, it benefits by the simplicity of the two-body decays. On the other, the spins of ${\bf B}_1$ and $\overline{{\bf B}}_2$ provide fruitful physical observables in experiments. 
Additionally, the production rate of $B$ is roughly three times larger than the one of the bottom baryons~\cite{ProductionRate}, making  $B\to {\bf B}_1 \overline{{\bf B}}_2$ an ideal place to probe CP violation with baryons. 
We note that the direct CP asymmetries in $B\to {\bf B}_1\overline{{\bf B}}_2$ have systematically estimated by Ref.~\cite{Chua:2022wmr} but the $B$ meson oscillation and  CP asymmetries in angular distributions have not been considered yet. 

The LHCb collaboration have measured the branching ratios of~\cite{LHCb:2016nbc,LHCb:2022lff},
\begin{eqnarray}\label{exp}
{\cal B} ( B^+ \to p \overline{\Lambda}) = \left(2.4_{-0.8}^{+1.0} \pm 0.3\right) \times 10^{-7}\,,
~~~{\cal B} ( B^0 \to p \overline{p}) = ( 1.27 \pm 0.14 ) \times 10 ^ { -8} \,,
\end{eqnarray}
and obtained a upper limit of~\cite{LHCb:2022lff} ,
\begin{equation}
{\cal B}(B_s^0 \to  p \overline{p}) < 5.1 \times 10 ^{-9}\,,
\end{equation}
at 95$\%$ confidence level, which is consistent with  ${\cal B}(B_s^0 \to  p \overline{p}) < 1 \times 10 ^{-10}$ predicted by  the chiral selection rule~\cite{Jin:2021onb,Chua:2022wmr}. 
In the theoretical studies, the calculated amplitudes depend heavily on QCD models, such as the pole model~\cite{Cheng:2001ub} and  sum rule~\cite{Chernyak:1990ag}. 
Nevertheless, in the literature ${\cal B} ( B^0 \to p \overline{p})$ is overestimated by an order. The reason behind it is closely related to the Fierz  identity~\cite{Cheng:2014qxa}.


This paper is organized as follows. In Sec. II, we define the decay parameters associated with spins and classify them in terms of CP-odd or CP-even parts. In Sec. III, we analyze the decay distributions of $B\to {\bf B}_1 \overline{{\bf B}}_2$. In Sec. IV, we estimate the decay parameters through the $^3 P_0$ model and chiral selection rule. Our numerical results are shown in Sec. V. Sec. VI is devoted to conclusions.

\section{Decay parameters}

In general, with $\hat{A}$ an arbitrary Hermitian operator, we can define a corresponding asymmetry as 
\begin{equation}~\label{asymmetry}
A\equiv \frac{
\Gamma(\lambda_A >0)  - \Gamma(\lambda_A < 0) 
}{\Gamma(\lambda_A >0)  + \Gamma(\lambda_A < 0) }\,,
\end{equation}
where $\Gamma$ represents the decay width, and $\lambda_A$ stands for the eigenvalue of $\hat{A}$. 
In decay final states, it is convenient to choose $\hat{A}$ as an $SO(3)$ rotational scalar, since the angular momenta are always constrained by the spins of the parent particles. In $B\to {\bf B}_1 \overline{{\bf B}}_2$, we simply have $J=0$ and the most simple operators are 
\begin{equation}\label{operators}
	\hat{\alpha} = \vec{s}_1 \cdot \hat{p} \,,\,\,\,\,\,\,\,\hat{\beta} = (\vec{s}_1 \times \vec{s}_2) \cdot \hat{p} \,,\,\,\,\,\,\,\,\hat{\gamma} = 2 \,\vec{s}_1 \cdot \vec{s}_2  \,,
\end{equation}
where $\vec{s}_1~(\vec{s}_2)$ is the spin operator of ${\bf B}_1~(\overline{{\bf B}}_2)$, and $\hat{p}$ is the 3-momentum norm of ${\bf B}_1$. 
From Eqs.~\eqref{asymmetry} and \eqref{operators}, we define 
{\small 
\begin{equation}
\alpha \equiv \frac{
	\Gamma(\lambda_\alpha >0)  - \Gamma(\lambda_\alpha < 0) 
}{\Gamma(\lambda_\alpha >0)  + \Gamma(\lambda_\alpha < 0) }\,, \,\,\,\,\,\beta \equiv \frac{
\Gamma(\lambda_\beta >0)  - \Gamma(\lambda_\beta < 0) 
}{\Gamma(\lambda_\beta >0)  + \Gamma(\lambda_\beta < 0) }\,,\,\,\,\,\,\gamma\equiv \frac{
\Gamma(\lambda_\gamma >0)  - \Gamma(\lambda_\gamma < 0) 
}{\Gamma(\lambda_\gamma>0)  + \Gamma(\lambda_\gamma < 0) }\,,
\end{equation}}
which affect the cascade decay distributions as we will demonstrate in the next section. Clearly, $\hat{\alpha}$ is a helicity operator and $\hat{\beta}$ describes the T-odd spin correlation.

Since  $\hat{\alpha}$ and $\hat{\beta}$ are both P-odd,  $\lambda_\alpha$ and $\lambda_\beta$ would flip signs  under the parity transformation. 
On the other hand, $\hat{\beta}$ is T-odd 
 and  $\beta$ is a  naively T-odd observable\footnote{
 It is not a truly T-odd observable as  $\hat{\beta}$ does not commute with the Hamiltonian. 
 }.
In general, a nonzero value of $\beta$ can be generated by the oscillations between $|\pm \lambda_\beta\rangle$ with the final state interactions. We utilize the CPT symmetry and define the true T-odd observables as 
\begin{equation}\label{betadefine}
\beta_w = (\beta\Gamma + \overline{\beta}\,\overline{\Gamma}) /(\Gamma+\overline{\Gamma}) \,,
\end{equation}
where overlines denote the charge conjugates. We emphasize that $\beta_w$ is clearly also a CP violating observable. The other CP violating observables can be defined to be  
\begin{equation}\label{CPdefine}
{\cal A}_{\text{dir}} =(\Gamma - \overline{\Gamma})/(\Gamma+ \overline{\Gamma}) \,, ~~~~ \alpha_w = (\alpha\Gamma + \overline{\alpha }\,\overline{\Gamma})/(\Gamma+ \overline{\Gamma}) \,, ~~~~ \gamma _w = (\gamma\Gamma - \overline{\gamma}\,\overline{\Gamma})/(\Gamma+ \overline{\Gamma})\,,
\end{equation} 
where the signs of Eqs.~\eqref{betadefine} and \eqref{CPdefine} are according to the parity of the responsible operators. To be explicit, $\hat{\alpha}$ and $\hat{\beta}$ are both P-odd, leading to the plus signs, whereas $\hat{\gamma}$ is P-even, resulting in the minus sign.
On the other hand, the CP-even observables are 
\begin{equation}
\beta_s =(\beta \Gamma -\overline{\beta}\,\overline{\Gamma})/(\Gamma+ \overline{\Gamma})\,,\quad\alpha_s = (\alpha\Gamma - \overline{\alpha}\,\overline{\Gamma})/(\Gamma+ \overline{\Gamma}) \,,\quad \gamma_s = (\gamma \Gamma + \overline{\gamma}\,\overline{\Gamma})/(\Gamma+ \overline{\Gamma}) \,,  
\end{equation}
 where the subscripts of $w$ and $s$ denote CP-odd and CP-even, respectively.

Clearly, to calculate the asymmetries in Eq.~\eqref{asymmetry}, it is necessary to obtain the eigenstates of $\hat{\alpha}, ~\hat{\beta}$ and $\hat{\gamma}$. To this end, we start with $\hat{\alpha}$ and express the others in the linear combinations of $|\lambda_\alpha\rangle$. With $J=0$ , the helicity states are given as 
\begin{equation}\label{helicities}
|\lambda_\alpha = \pm 1/2 \rangle  = \int^{2 \pi}_ 0\int^1_{-1} R_z(\phi) R_y(\theta) \left|\hat{p}=\hat{z}, \lambda_1 =  \lambda_2 = \pm 1/2 \right\rangle d \cos\theta d \phi \,,
\end{equation}
where $\lambda_{1(2)}$ is the helicity of ${\bf B}_1~(\overline{{\bf B}}_2)$, and $R_i$ is the rotational operator pointing toward the $i$th direction.
Note that the operators in Eq.~\eqref{operators} commute with $R_i$ and~\cite{BVV} 
\begin{equation}\label{rule}
(s_j)_{\pm} | \hat{p}=\hat{z}, \lambda_j = \mp 1/2\rangle = |\hat{p}=\hat{z}, \lambda_j = \pm 1/2\rangle\,\,\,\,\,\,\,\,\text{for}~~j \in \{1,2 \}\,, 
\end{equation}
with $(s_j)_\pm  = (s_j)_x \pm i (s_j)_y$. Together with Eqs.~\eqref{operators}, \eqref{helicities} and \eqref{rule}, the  eigenstates of $\hat{\beta}$ and $\hat{\gamma}$ are then given as 
\begin{eqnarray}\label{eigen}
|\lambda_\beta =\pm 1/2 \rangle &=& \frac{1}{\sqrt{2}}\left(
|\lambda_\alpha = 1/2\rangle \mp i | \lambda_\alpha = -1/2\rangle
\right)\,,\nonumber\\
|\lambda_\gamma =\pm 1/2 \rangle &=& \frac{1}{\sqrt{2}}\left(
|\lambda_\alpha = 1/2\rangle \pm  | \lambda_\alpha = -1/2\rangle
\right)\,.
\end{eqnarray} 
The asymmetry parameters defined in Eq.~\eqref{asymmetry} are then given as
\begin{equation}\label{parameters}
\alpha  = \frac{|H_+|^2 - |H_-|^2}{|H_+|^2 + |H_-|^2}\,,\,\,\,\,\,\,\,\beta = \frac{2\Im(H_-^*H_+)}{|H_+|^2 + |H_-|^2}\,,\,\,\,\,\,\,\,\gamma   = \frac{2\Re( H_-^* H_+ ) }{|H_+|^2 + |H_-|^2}\,,
\end{equation}
with the identity 
\begin{equation}\label{equality of parameter}
  1-\left(\alpha^2 + \beta ^2  + \gamma ^2\right) =0\,. 
\end{equation}
 Here,  $H_\pm$ stand for the helicity amplitudes, defined by 
 \begin{equation}\label{20}
H_\pm = \langle \lambda_\alpha = \pm 1/2| i{\cal H}_{eff} | B\rangle = \sum_jH^{\pm}_j  \exp \left(i\phi_{j_s}^\pm  +i\phi_{j_ w}^\pm \right) \,,
 \end{equation}
where ${\cal H}_{eff}$ is the effective Hamiltonian,
$H_j^\pm$ are real, and $\phi^{j_s}_{\pm }$ and $\phi^{j_w}_{\pm}$ are the strong and the weak CP phases, respectively.
The charge conjugate ones can be obtained by taking the CP transformation
 \begin{equation}\label{21}
 \overline{H}_\mp = \langle \lambda_\alpha = \mp 1/2| i{\cal H}_{eff} | \overline{B}\rangle = \sum_j H^{\pm}_j  \exp \left(i\phi_{j_s}^\pm  - i\phi_{j_ w}^\pm \right) \,.
\end{equation}
 Notice that $\lambda_ \alpha$ flips sign after the CP transformation.

For $B_{(s)}^0$, the decay parameters in Eq.~\eqref{parameters} depend on time due to the $B-\overline{B}$ mixing,  given  by 
 \begin{eqnarray}\label{oscillation}
&&|B^0_{(s)} (t) \rangle = g_+(t) |B_{(s)}^0 (t=0)\rangle + \frac{q}{p}g_-(t) | \overline{B}_{(s)}^0 (t=0)\rangle\,,\nonumber\\
&&|\overline{B}^0_{(s)} (t) \rangle =\frac{p}{q} g_-(t) |B_{(s)}^0 (t=0)\rangle + g_+(t) | \overline{B}_{(s)}^0 (t=0)\rangle\,,
 \end{eqnarray}
 where $p $ and $ q$  are the mixing parameters, and 
 \begin{equation}\label{gpm}
g_+(t) \pm g_- (t) =  e^{\left(
-\Gamma  \mp \Delta \Gamma   /2 
 \right)t / 2  }
e^{-i \left( M \pm \Delta M/2
 \right) t }\,.
 \end{equation}
The parameters associated with the masses and the decay widths  are defined as
\begin{eqnarray}
&&M = (M_{H} + M_{L})/2\,, \quad \Gamma = (\Gamma_H + \Gamma_L)/2\,,\nonumber\\
&& M_\Delta = M_H - M_L \,,\quad  \Gamma_\Delta =  \Gamma_L - \Gamma_H  \,,
\end{eqnarray}
where $m_{L,H}$ and $\Gamma_{L,H}$ are the masses and  total decay widths of the light and heavy neutral $B$ mesons, respectively.
Clearly,  $g_\pm(t)$, $M_{H,L}$ and  $\Gamma_{H,L}$  depend on whether $B^0$ or $B^0_s$ is in question. We do not explicitly show the dependence as long as it does not cause confusion.
In this work, we take $q=p$, corresponding to that CP is conserved in the oscillations, which  causes some errors at the $O(10^{-3})$ level.  In the future studies, the approximation is not necessary if a higher precision is desired.

Using Eqs.~\eqref{20} and \eqref{21}, we arrive at 
\begin{eqnarray}
&&\langle \lambda_\alpha =\pm  1/2 |i {\cal H}_{eff} | B^0_{(s)}(t)\rangle = g_+  (t)H_{\pm}+ g_- (t) \overline{H}_\pm\,,\nonumber\\
&&\langle \lambda_\alpha =\pm  1/2 |i {\cal H}_{eff} | \overline{B}^0_{(s)}(t)\rangle = g_+(t) \overline{H}_\pm + g_-  (t) H_\pm\,,
\end{eqnarray}
leading to 
\begin{eqnarray}\label{20}
D(t) &=& (|g_+|^2 + |g_-^2| ) D + 4 \Re(g_+g_-^*) \left(
\Re (H_+\overline{H}_+^*  +  H_ -\overline{H}_-^*)
\right)\nonumber\\
{\cal A}_{\text{dir}}(t) &=& \left[ (|g_+|^2- |g_-|^2) {\cal A}_{\text{dir}}(0) D + 4 \Im (g_+ g_-^* ) \left(
\Im (H_+\overline{H}_+^* + H_ -\overline{H}_-^*)
\right) \right] /D (t) \,,\nonumber \\
\alpha_s(t)  &=& \left[ (|g_+|^2- |g_-|^2) \alpha_s(0) D + 4 \Im (g_- g_+^* ) \left(
\Im (H_+\overline{H}_+^* -  H_ -\overline{H}_-^*)
\right) \right] /D (t) \,, \nonumber\\
\alpha_w(t) &=& \left[ (|g_+|^2 + |g_-|^2) \alpha_w(0) D + 4 \Re (g_+ g_-^* ) \left(
\Re (H_+\overline{H}_+^*  - H_ -\overline{H}_-^*)
\right)\right]   /D (t) \,,\nonumber\\
\beta_s(t)  &=& \left[ (|g_+|^2- |g_-|^2) \beta_s(0) D + 4 \Im ( g_+^*g_- ) \left(
\Re (H_-^*\overline{H}_+ -  H_+\overline{H}_-^*)
\right) \right] /D (t) \,, \nonumber\\
\beta_w(t) &=& \left[ (|g_+|^2 + |g_-|^2) \beta_w(0) D + 4 \Re (g_+^*  g_-) \left(
\Im (H_-^*\overline{H}_+  + H_+\overline{H}_-^*)
\right)\right]   /D (t) \,,\nonumber\\
\gamma_s(t)  &=& \left[ (|g_+|^2+ |g_-|^2) \gamma_s(0) D + 4 \Re (g_- g_+^* ) \left(
\Re (H_-^*\overline{H}_+ +  H_+\overline{H}_-^*)
\right) \right] /D (t) \,, \nonumber\\
\gamma_w(t) &=& \left[ (|g_+|^2 - |g_-|^2) \gamma_w(0) D - 4 \Im (g_- g_+^*) \left(
\Im (H_-^*\overline{H}_+  - H_+\overline{H}_-^*)
\right)\right]   /D (t) \,,
\end{eqnarray}
where the denominator is $D = |H_+|^2+|H_-|^2+|\overline{H}_+|^2+|\overline{H}_-|^2$. 
In the experiments, the measured quantities correspond to the ones averaged from $t_1$ to $t_2$. By taking $t_1= 0 $ and $t_2\to \infty$, we find that 
\begin{eqnarray}\label{21}
&&\langle
|g_+|^2 + |g_-|^2 \rangle 
= \frac{1}{\Gamma} \frac{4}{4 -x ^2 }
\,,
\qquad
\langle
|g_+|^2 - |g_-|^2 \rangle = \frac{1}{\Gamma} \frac{1}{1 +y ^2 }\,,
\nonumber\\
&&\langle g_+ (t)g_- ^*(t) \rangle
=
\frac{1}{\Gamma} \left(  \frac{- x }{4-x^2 }
- \frac{i y}{1+ y }
\right) \,,
\end{eqnarray}
with $(x,y) = ( \Gamma_\Delta / \Gamma , M_\Delta / \Gamma )$. 
In this work, 
the values of the oscillating parameters are from the Particle Data Group (PDG)~\cite{pdg}, given by
\begin{eqnarray}\label{22}
(x,y)_{B_0}=(0.001,0.77), &(x,y)_{B_s^0} =(0.128,27)  \,.
\end{eqnarray}

\section{ angular distributions}
From Eq.~\eqref{helicities},  the decay distributions of $B\to {\bf B}_1 \overline{{\bf B}}_2$ are trivial since $B$ is spinless. Hence, the asymmetry parameters defined in Eq.~\eqref{parameters} are essentially described by the spin correlations between ${\bf B}_1$ and $\overline{{\bf B}}_2$. In the LHCb experiments, spins can not be measured directly and we therefore shall seek the spin effects  in their cascade decays. 

If ${\bf B}_1\in \{
\Xi^{0,-}, \Sigma^\pm, \Lambda
\}$,
it would
 consequently decay to an octet baryon~$({\bf B}_1')$ and a pion, while $H_\pm $  interfere via the cascade decays.  The partial decay widths are  proportional to 
\begin{equation}
\frac{\partial \Gamma}{\partial \cos \theta_1 }  \propto \sum_{\lambda_1}\left|
\sum_\lambda H_\lambda A_{1,\lambda_1}  d^{\frac{1}{2}}(\theta_1) ^\lambda\, _{\lambda_1} 
\right|^2\,,
\end{equation}
 with $\theta_1$  defined as the polar angles in the ${\bf B}_1$ helicity frame~(see FIG.~1), 
 resulting in
\begin{eqnarray}
&&{\cal D}_1(\theta_1) \equiv\frac{1}{\Gamma} \frac{\partial \Gamma}{\partial \cos \theta_1 }  =\frac{1}{2}\left(
1+ \alpha \alpha_1 \cos \theta_1 
\right)\,,
\end{eqnarray}
where we have used $\left|A_{1, \pm} \right|^2 = (1\pm \alpha_{1})/2 $ in the last equation with $\alpha_1$ the up-down asymmetry parameter for ${\bf B}_1 \to {\bf B}_1' \pi $, and $d^{J}(\theta) ^M \, _N$ is the  Wigner-$d$ matrix, 
 defined by $d^J(\theta)^M\,_N\equiv \langle J, M | R_y(\theta) | J, N \rangle$.  
On the other hand, if  $\overline{{\bf B}}_2\in \{
\overline{\Xi}^{0,+}, \overline{\Sigma}^\pm , \overline{\Lambda}
\}$, it would sequentially decay to $\overline{{\bf B}}_2'\pi $, and   the angular distributions are given as 
\begin{eqnarray}
&&{\cal D}_2(\theta _2) \equiv\frac{1}{\Gamma} \frac{\partial^2\Gamma}{\partial \cos \theta_2\partial \phi_2}  =\frac{1}{4\pi}\left(
1+ \alpha \overline{\alpha}_2 \cos \theta_2
\right)\,,
\end{eqnarray}
where $ \overline{\alpha}_2$ is the up-down asymmetry parameters for $\overline{{\bf B}}_2 \to \overline{{\bf B}}_2' \pi $.
Here, $\theta_{1,2}$ are defined as the angles between $\vec{p}_{1,2}$ and $\vec{p}_{1,2}^{\,\prime} $ with $\vec{p}_{1,2}^{\,(\prime)}$ the 3-momentum of ${\bf B}_{1,2}^{(\prime)}$; see FIG.~1 with $B_s^0 \to \Lambda \overline{\Lambda}$ for illustration.

\begin{figure}
	\includegraphics[width=.6\linewidth]{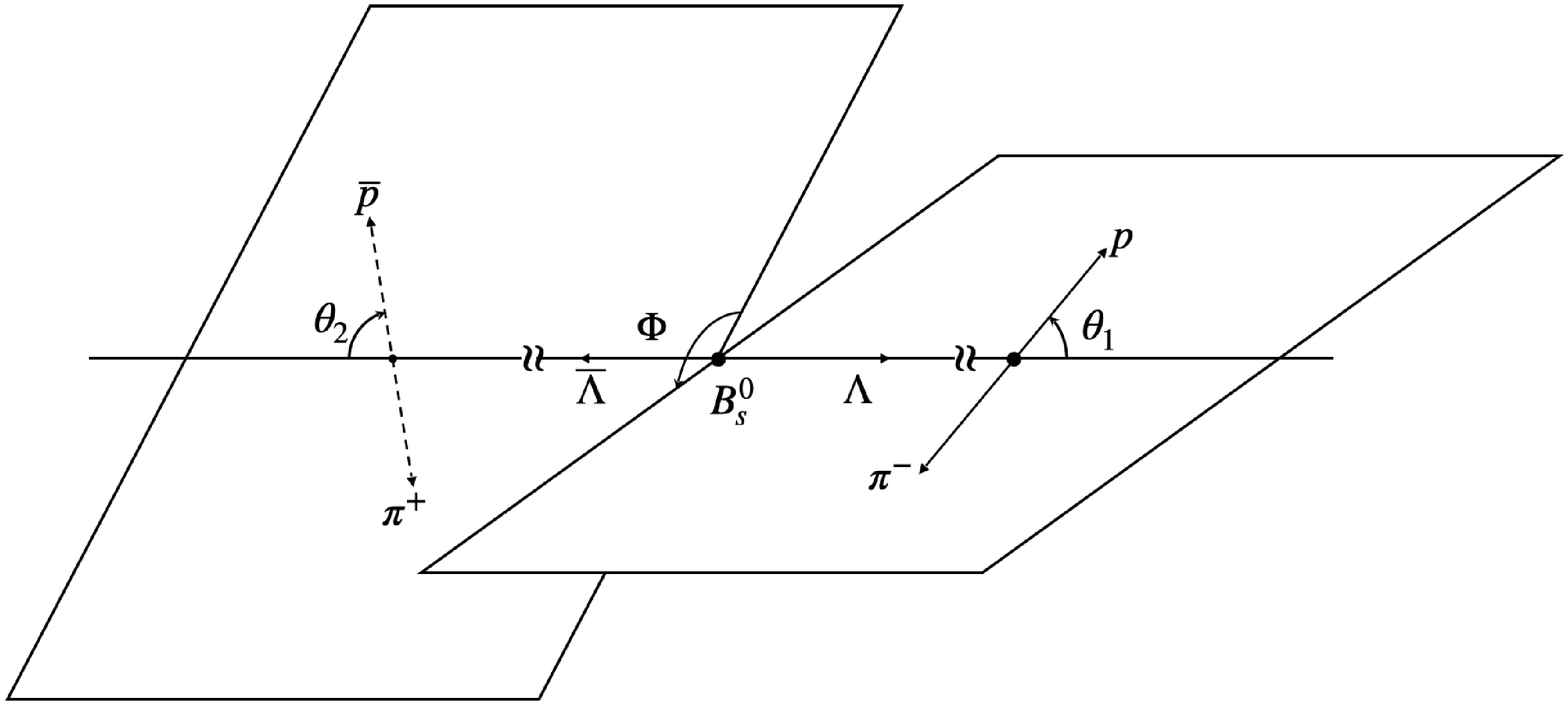}  
\caption{
	Angles in ${\cal D}(\vec{ \Omega})$ for $B_s^0 \to \Lambda (\to p \pi^-) \overline{\Lambda} (\to \overline{p}\pi ^+) $.
}
\end{figure}

When ${\bf B}_1$ and $\overline{{\bf B}}_2$ both decay subsequently, there would be three independent 3-momenta in the final states and it is possible to define an azimuthal angle. The 
angular distributions
for 
$ B \to {\bf B}_1(\to {\bf B}_1' \pi ) \overline{{\bf B}}_2(\to \overline{{\bf B}}'_2 \pi ) $ are given as
	\begin{eqnarray}
{\cal D}({\vec{ \Omega}})&\equiv&	\frac{1}{\Gamma}\frac{\partial^3\Gamma}{
			\partial \Phi  \partial \cos \theta_1 \cos \theta_2}	\nonumber\\
		&=&\frac{1}{8\pi }\left[
		1+\alpha_1\overline{\alpha}_2\cos \theta_1 \cos \theta_2+
		\alpha (\alpha_1\cos \theta_1+\overline{\alpha}_2\cos \theta_2 )\right.\nonumber\\
&&		\left.+
		2 \alpha_1 \overline{\alpha}_2 \sin\theta_1 \sin \theta_2\left(
		\gamma \cos \Phi - \beta \sin \Phi  
		\right)\right],\,
	\end{eqnarray}
where $\theta_{1,2}$ are defined as same as the previous ones. By integrating $\theta_{1,2}$, we find
\begin{equation}
\frac{1}{\Gamma}\frac{\partial \Gamma}{\partial \Phi} \propto 1+ \frac{\pi^2 }{8}\alpha_1 \alpha_2 (\gamma \cos\Phi - \beta \sin \Phi ) \,.
\end{equation}
We see that $\gamma$ and $\beta$ are shown in the double cascade decays  and  possible to be measured in experiments. It is straightforward to see
 that by integrating $\phi_{1(2)}$ and $\theta_{1(2)}$, ${\cal D}(\vec{ \Omega})$ reduces to ${\cal D}_{1(2)}(\theta_{1(2)})$ as expected.

\section{Model calculations}

\begin{figure}[t!]
\centering
\includegraphics[width=0.33\textwidth]{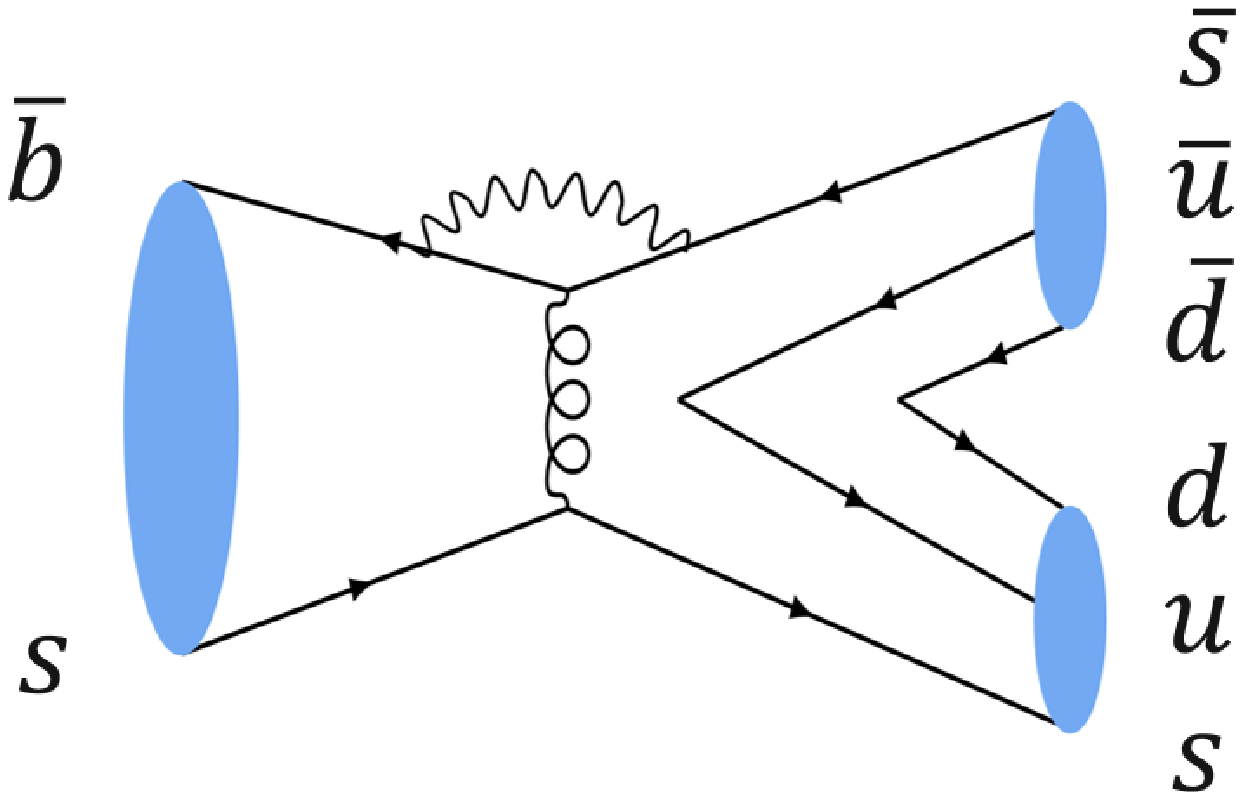}\hspace{1cm}
\includegraphics[width=0.33\textwidth]{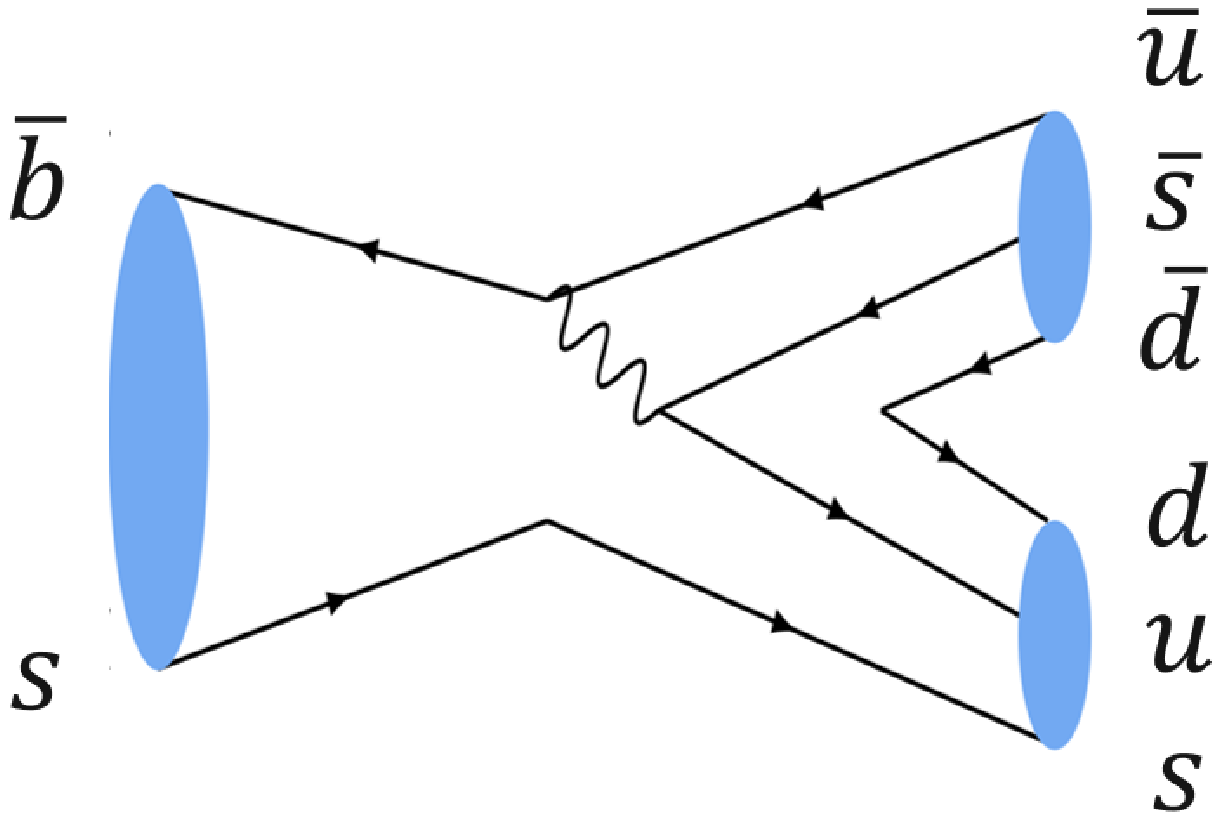}
\caption{Quark diagrams for $B_s^0 \to \Lambda \overline{\Lambda}$. }
\label{JpsiFeyn}
\end{figure}

In this sections  we  estimate the decay parameters in the SM through model calculations.
In addition to $B^0 \to p\overline p$ and $B^+ \to p \overline{\Lambda}$, we also consider the processes of $B_s^0 \to \Lambda \overline{\Lambda}$ and $B_s^0 \to \Sigma^+ \overline{\Sigma}^-$, which are promising to be measured in the near future.
The  considered quark diagrams are shown in FIG.~\ref{JpsiFeyn}, where the left one is factorizable, whereas the right one is
not, denoted by ${\cal A}_P$ and ${\cal A}_T$, respectively.
For the $b\to f$ transition  at the quark level, 
we have 
\begin{eqnarray}
&&{\cal A}_P=
\zeta \langle 
{\bf B}_1 ;\overline{{\bf B}}_2  | \overline{q}_B  (1+\gamma_5)  f  |0 \rangle
=\zeta\overline{u}_1 
\left( F(q^2) + F_5 (q^2) \gamma_5
\right) 
v_2\,,\nonumber\\
&&\zeta = \frac{G_F}{\sqrt{2}} V_{t b}^* V_{t f} f_B\left(c_6+c_5 / 3\right) \frac{2 m_B^2}{m_b}
\end{eqnarray}
where  $q_B =(u,d,s)$ is the light flavor of the $B$ meson,
$G_F$ corresponds to the Fermi constant,  $V$ is the Cabibbo–Kobayashi–Maskawa matrix, $f_B$ ($m_B$) is the decay constant (mass) of $B$, $m_b\approx 4.5$ GeV is the $b$ quark mass,  $u_1$ ($v_2$) 
is the Dirac spinor of ${\bf B}_1 $  ($\overline{{\bf B}}_2$), and $F_{(5)}(q^2)$ is the (pseudo)scalar form factor. 
In our previous work~\cite{Jin:2021onb}, we have used the crossing symmetry and analytical continuations to calculate $F_{(5)}(q^2)$.  It was done by assuming that the singularity does not exist in the $q^2$ dependence of  form factors.  

In this work,
we adopt the $^3P_0$ model to calculate the form factors. 
The model asserts that creations of valence quarks can be approximated by scalar operators at the limit of hadrons being at rest; see Refs.~\cite{Micu:1968mk,LeYaouanc:1972vsx,Ackleh:1996yt,Segovia:2012cd,Simonov:2011cm} for instance.  
To accommodate the fact that ${\bf B}_1$ and $\overline{ {\bf B}} _2$ in $B$ meson decays are far from at rest, we calculate the form factors at $\vec{v}=0$ and take their  dependencies on $\vec{v}$ as ~\cite{BESIII:2019hdp}
\begin{eqnarray}
F_{(5)} (\gamma)  &=&
\frac{{\cal F}_{(5)}}{1-5.29\gamma+7.07 \gamma^2-0.31 \gamma^3+0.41 \gamma^4}  \,,
\end{eqnarray}
where $\gamma = 1/\sqrt{1-v^2}$, $\vec{v}$ is the velocity of ${\bf B}_1$ at the Briet frame, ${\cal F}_{(5)}$ is a constant to be fitted at $\gamma = 1$ from the $^3P_0$ model, and the coefficients of $(5.29,7.07,-5.29,0.41)$ are  extracted from the experiments of  $e^-e^+ \to p \overline{p}$.

The method of  calculating ${\cal F}_{(5)}$ is similar to the one used in  $J/\psi \to \Lambda \overline{\Sigma}^0$~\cite{Jpsi} and  we briefly quote the formalism used in the numerical evaluations here.
We define the amplitudes  
\begin{eqnarray}
A^{{\bf B}_1\overline{ {\bf B}}_2}  &\equiv& \zeta \langle 
{\bf B}_1,   \uparrow  ; \overline{{\bf B}}_2,   \uparrow  | \overline{q}_B   f  |0 \rangle  \,,
\nonumber\\
A^{{\bf B}_1\overline{ {\bf B}}_2}_5  &\equiv& \zeta  \langle 
{\bf B}_1,   \uparrow  ; \overline{{\bf B}}_2,   \uparrow  | \overline{q}_B  \gamma_5  f  |0 \rangle 
\,.
\end{eqnarray}
The factorizable parts of the helicity amplitudes are then given by 
$H^{fac}_\pm = A^{{\bf B}_1 \overline{{\bf B}}_2}
\pm A_5^{{\bf B}_1 \overline{{\bf B}}_2}$. 
For $B^+ \to p \overline{\Lambda}$ as an example, we find~\cite{Liu:2022pdk}   
\begin{eqnarray}
A^{p \overline{\Lambda}}_{(5)}   &=&
\gamma \zeta \gamma_q^2  \frac{N^2  }{2} \int d^3\vec{x}_\Delta \Gamma^{(5)} _{  \uparrow \downarrow  }
\Big[ 
\big(
2 E^{u}_{\uparrow\uparrow} E^{d}_{\downarrow\downarrow}
+ E^{u}_{\downarrow\downarrow}E^{d}_{\uparrow\uparrow}
- 2E^{u}_{\uparrow\downarrow}E^{d}_{\downarrow\uparrow}  \nonumber\\
&& 
- E^{u}_{\downarrow \uparrow} E^{d}_{\uparrow \downarrow }
+\Gamma^{(5)} _{ \uparrow  \uparrow  }  
\left(
E^{u}_{\downarrow\uparrow} E^{d}_{\downarrow\downarrow}
-E^{d}_{\downarrow\downarrow} E^{u}_{\downarrow\uparrow}
\right) 
\Big]\,,
\end{eqnarray}
where
$N$ is the normalization constant, 
the $\vec{x}_\Delta$ dependences of 
$E(\vec{x}_\Delta)$ and $\Gamma(\vec{x}_\Delta)$ have not been written out explicitly, and $\gamma_q \approx 0.3 $ is the strength of the quark production in the $3P^0$ model. 
Here, $\Gamma_{\lambda_1\lambda_2}$ represents the overlapping of the quarks which participate in the weak interactions, whereas $E^q_{\lambda_1\lambda_2}$ of the spectator quark.  
For $\vec{v} =0 $, the formalism is reduced to 
\begin{eqnarray}
&&\Gamma_{\uparrow\downarrow} ( \vec{x}_\Delta) 
= E_{\uparrow\downarrow} ( \vec{x}_\Delta)    \,,
~~~\Gamma_{\uparrow\uparrow} ( \vec{x}_\Delta)
= E_{\uparrow\uparrow} ( \vec{x}_\Delta) \,,~~~
\Gamma^5_{\uparrow\uparrow} ( \vec{x}_\Delta) 
= 0\nonumber\,, \\
&&\Gamma^5_{\uparrow\downarrow} ( \vec{x}_\Delta) 
= -2 \pi \int \rho d\rho dz  \left[ u_+ u _- + v_+v_- (\rho^2 +z_+z_-) \right]\,,
\end{eqnarray}
where the explicit forms of $N$, $E(\vec{x}_\Delta)$, $u_\pm$, $v_\pm$ and $z_\pm$ and the computing techniques can be found in  Ref.~\cite{Jpsi}. Likewise, the expressions for $B_0 \to p\overline{p}$, $B^0_{s} \to \Lambda\overline{\Lambda}$ and $B^0_s \to \Sigma^+ \overline{\Sigma}^-$  are
\begin{eqnarray}
A^{p \overline p}_{(5)} &=& 
 \zeta \gamma_q^2  \frac{N^2  }{2} \int d^3\vec{x}_\Delta
(
4E^{u} _{  \downarrow\uparrow  } E^{u} _{  \uparrow\uparrow  } \Gamma ^{(5)} _{ \downarrow \downarrow  } + 4E^{u} _{  \downarrow\uparrow  } E^{u} _{  \downarrow\downarrow  } \Gamma ^{(5)} _{  \uparrow\uparrow  }   \nonumber\\
&&
- 4E^{u} _{  \downarrow\uparrow  }E^{u} _{  \downarrow\uparrow  }\Gamma^{(5)} _{\uparrow   \downarrow } -2\Gamma^{(5)} _{  \downarrow\uparrow  }(E^{u} _{  \downarrow\uparrow  } E^{u} _{  \uparrow\downarrow  } +E^{u} _{  \uparrow\uparrow  }E^{u} _{  \downarrow\downarrow  } )) \,,\nonumber\\
A_{(5)}^{\Lambda \overline{\Lambda} }   &=&
 \zeta \gamma_q^2  \frac{N^2  }{2} \int d^3\vec{x}_\Delta \Gamma^{(5)} _{  \downarrow\uparrow  }
\big(
E^{u}_{\uparrow\uparrow} E^{d}_{\downarrow\downarrow}
+ E^{u}_{\downarrow\downarrow}E^{d}_{\uparrow\uparrow})\,, \nonumber\\
A_{(5)}^{\Sigma^+ \overline{\Sigma}^-} &=& 
 \zeta \gamma_q^2  \frac{N^2  }{2} \int d^3\vec{x}_\Delta
(
4E^{u} _{  \downarrow\uparrow  } E^{u} _{  \uparrow\uparrow  } \Gamma ^{(5)} _{ \downarrow \downarrow  } + 4E^{u} _{  \downarrow\uparrow  } E^{u} _{  \downarrow\downarrow  } \Gamma ^{(5)} _{  \uparrow\uparrow  }   \nonumber\\
&&
- 4E^{u} _{  \downarrow\uparrow  }E^{u} _{  \downarrow\uparrow  }\Gamma^{(5)} _{\uparrow   \downarrow } -2\Gamma^{(5)} _{  \downarrow\uparrow  }(E^{u} _{  \downarrow\uparrow  } E^{u} _{  \uparrow\downarrow  } +E^{u} _{  \uparrow\uparrow  }E^{u} _{  \downarrow\downarrow  } ))\,.
\end{eqnarray}
We note that we have assumed the $SU(3)_F$ symmetry to simplify the formalism.

With the factorizable amplitudes, we find\footnote{
We note that the results are consistent with the use of the crossing symmetry, where we found that ${\cal B}_{fac}(B^+ \to p\overline{\Lambda}) = (1.3 \pm 0.1) \times 10 ^{-7}$, ${\cal B}_{fac}(B_s^0 \to \Lambda \overline{\Lambda}) = (1.6 \pm 0.1) \times 10 ^{-7}$, ${\cal B}_{fac}(B_s^0 \to \Sigma^+ \overline{\Sigma}^-) = (2.9 \pm 0.2) \times 10 ^{-7}$ and ${\cal B}_{fac} (B^0 \to p\overline{p}) = 0.2 \times 10 ^{-8}$~\cite{Jin:2021onb}. The numerical results in this work serve as an illustration for the CP violating quantities without the discussions of the error analyses.
}
\begin{eqnarray}\label{Ba}
&&{\cal B}_{fac}(B^+ \to p \overline{\Lambda} ) = 1.57 \times 10 ^{-7}\,, ~~
{\cal B}_{fac} (B_s^0 \to \Lambda \overline{\Lambda}) = 2.6\times 10^{-7}\,,\nonumber\\&&{\cal B}_{fac} (B_s^0 \to \Sigma^+ \overline{\Sigma}^-) = 2.17\times 10^{-7}\,, ~~{\cal B}_{fac} (B^0 \to p \overline{p} ) = 0.21 \times 10^{-8}\,,
\end{eqnarray}
where
the subscript in ${\cal B}_{fac}$ indicates that only the factorizable part of the amplitude is considered. In Eq.~\eqref{Ba},  
${\cal B}_{fac}(B^+ \to p \overline{\Lambda} ) $ is compatible with the experimental data in Eq.~\eqref{exp} but ${\cal B}_{fac}(B^0 \to p \overline{p} ) $ is 6 times smaller.
The reason can be traced back to that ${\cal A}_T$ is considered yet.
For $|V_{ub}^* V_{ud}|\gg |V_{ub}^* V_{us}|$ and $|V_{tb}^* V_{td}|\ll |V_{tb}^* V_{ts}|$\,, ${\cal A}_P$ and ${\cal A}_T$ play a leading role in the $b\to s$ and $b\to d$ transitions, respectively. 

For the nonfactorizable diagram in the right hand side of FIG. 2, we fit its amplitude from the experimental branching ratios in Eq.~\eqref{exp}. For an estimation, we assume the relative complex phase between two diagrams to be maximum\footnote{The cases with vanishing relative strong phases
are studied in Ref.~\cite{Jin:2021onb}, which would lead to zero $A_{dir}$. 
}. The ratios of the nonfactorizable amplitudes among the channels are determined by the chiral selection rule described in Ref.~\cite{Jin:2021onb}.
 The ones relevant to this work read
\begin{equation}
{\cal A}_{T}(B^0 \to p\overline{p}) :
{\cal A}_{T}(B^+ \to p\overline{\Lambda}) : 
{\cal A}_{T}(B_s^0 \to \Sigma^+ \overline{\Sigma}^-) : 
{\cal A}_{T}(B_s^0 \to \Lambda \overline{\Lambda})
= 2 : -\sqrt{6} : 2 : -3 \,. 
\end{equation}
 The numerical results at $t=0$ and the time-averaged ones are given in Tables~\ref{t1} and \ref{t2}, respectively, where
${\cal B}_{av}$ stands for the CP-even part of the branching ratio and  is unaffected by the $B$ meson oscillations.

\begin{table}
\begin{center}
	\caption{
 The decay parameters at $t=0$, where the 
branching ratios  are 
 in units of $10^{-7}$ and the other in units of $10^{-2}$. The upper and lower columns correspond to the results with positive and negative strong phases, respectively.
	}\label{t1}
\begin{tabular} {lrrrrrrrrrrrrrrrrrrr}
		\hline
		\hline
channels &${\cal B}_{av} $ ~&~$A_{dir}$~&~$\alpha_s $~&~$ \beta_s$~&~$  \gamma_s$~&~$\alpha_w $~&~$ \beta_w$~&~$  \gamma_w$\\
		\hline
  \multirow{2}{*}{$B^0 \to p \overline{p} $}&$0.127$&$29.1$&$ -93.8 $&$ 14.7 $&$24.3 $&$-29.1 $&$-12.8$&$ 15.2 $\\
&$0.127$&$-40.3$&$ -93.8 $&$ -20.4 $&$- 5.3 $&$ 40.3 $&$17.8$&$ -21.0 $\\
\hline
  \multirow{2}{*}{$B^+ \to p \overline{\Lambda} $}&$2.31$&$ -3.1$&$62.7 $&$ 6.2 $&$ -77.3 $&$3.1$&$-1.5$&$ 6.4 $\\
&$2.31$&$ 4.2$&$59.7$&$ -8.4 $&$ -79.3 $&$ -4.2 $&$ 2.0$&$ -8.7 $\\
\hline
   \multirow{2}{*}{$B_s^0 \to  \Sigma^+ \overline{\Sigma}^- $}&$4.48$&$2.8$&$ -1.1 $&$- 2.7 $&$- 99.9$&$-2.8$&$ 0.7$&$-2.8$\\
   &$4.42$&$-3.9$&$ 0.3 $&$3.9$&$- 99.8$&$3.9$&$  -0.9$&$ 3.9 $\\
\hline
   \multirow{2}{*}{$B_s^0 \to \Lambda \overline{\Lambda} $}&$7.42$&$3.3 $&$ -1.2$&$ -3.2 $&$ -99.9 $&$ -3.3 $&$0.8$&$ -3.3 $\\
&$7.29$&$-4.7 $&$ 0.5$&$ 4.5 $&$ -99.8 $&$4.7 $&$ -1.1$&$ 4.7 $\\
		\hline
  \hline
	\end{tabular}
\end{center}
\end{table}

\begin{table}
\begin{center}
	\caption{
 The time-averaged decay parameters, where the 
branching ratios  are 
 in units of $10^{-7}$ and the other in units of $10^{-2}$. 
 The upper and lower columns correspond to the results with positive and negative strong phases, respectively.
	}\label{t2}
\begin{tabular} {l rrrrrrrrrrr}
		\hline
  \hline
channels &${\cal B} _{av} $~&~$\langle A_{dir} \rangle$~&~$\langle\alpha_s\rangle $~&~$\langle \beta_s \rangle$~&~$  \langle\gamma_s \rangle $~&~$\langle \alpha_w \rangle $~&~$ \langle \beta_w \rangle $~&~$   \langle \gamma_w \rangle$\\
				\hline
		\multirow{2}{*}{$B^0 \to p \overline{p} $}&$ 0.127$&$ 26.2$&$-82.9 $&$13.2$&$ 25.1 $&$-29.5$&$ -13.0$&$13.6$\\
  &$ 0.127$&$-36.9$&$  -82.8$&$ -18.7$&$ -5.1$&$ 41.7$&$ 18.4$&$-19.3$\\
  \hline  
		\multirow{2}{*}{$B_s^0 \to  \Sigma^+ \overline{\Sigma}^- $}&$ 4.48$&0.0&0.0&0.0&$-99.9$&$-2.8$&$0.7$&0.0\\ 
   &$ 4.42$&0.0&0.0&0.0&$-99.8$&$3.9$&$-0.9$&0.0\\   
          \hline 
		\multirow{2}{*}{$B_s^0 \to \Lambda \overline{\Lambda} $}&$ 7.42$&0.0&0.0&$ 0.0 $&$ -99.9$&$ -2.2 $&$0.5$&0.0\\
 &$ 7.29$&0.0&0.0&0.0&$ -99.9$&$ 3.1 $&$ -0.7 $&0.0\\     
		\hline
		\hline
	\end{tabular}
\end{center}
\end{table}

We note there remain ambiguities in strong phases related by complex conjugates. The two possibilities would lead to an opposite sign in $A_{dir}$ and may be determined by the future experiments. In the tables, two scenarios are  presented in the upper and lower columns, respectively. 
 For the $b\to d $ transition, namely $B^0\to p \overline{p}$, $A_{dir}$ at $t=0$ is found to be as large as  $-40.3\%$ or $29.1\%$, whereas $|A_{dir}|<7\%$ in general for the $b\to s$ transition, namely $B^+\to p \overline{\Lambda}$ and $B_s^0 \to \Sigma^+ \overline{\Sigma}^-, \Lambda \overline{\Lambda}$.  More importantly, due to the violent oscillation (see Eq.~\eqref{22})  between $B_s^0$ and $\overline{B}_s^0$, the differences between 
the $B_s^0$ and $\overline{B}_s^0$ decays in 
the branching ratios are washed out quickly, leading to tiny $\langle A_{dir}\rangle$. In contrast, the $B^0$ oscillation is much more gentle, and we have $A_{dir}(t=0) \approx \langle A_{dir}\rangle $. This argument is supported by the explicit calculations as well as Eqs.~\eqref{20} and \eqref{21}. 
The same suppression due to the oscillation is  found  also in $\langle\gamma_w\rangle$ but not $\langle\alpha_w\rangle$ and $\langle\beta_w\rangle$. We conclude that $\langle\alpha_w\rangle$ and $\langle\beta_w\rangle$ are ideal observables to be probed in the experiments. In particular,  $(\langle \alpha_w\rangle, \langle \beta_w\rangle ) $ for $B_s^0 \to \Lambda \overline{\Lambda}$ is estimated to be $(-2.2\%, 0.5\%)$ or $(3.1\%, -0.7\%) $, and the relative large branching ratio of $B_s^0 \to \Lambda \overline{\Lambda}  $ would  benefit the experimental measurement.  


\section{conclusions}

We have studied the decay observables in $B^0\to p \overline{p}$, $B^+ \to p \overline{\Lambda} (\to \overline{p} \pi^+)$, $B_s^0\to \Sigma ^+ (\to p\pi^0 ) \overline{\Sigma}^-(\to \overline{p} \pi^0 )$ and $B_s^0 \to \Lambda(\to p \pi^-) \overline{\Lambda}(\to \overline{p} \pi^+)$.
The spin-related CP-odd and CP-even quantities in $B\to {\bf B}_1 \overline{{\bf B}}_2$
have been examined. 
Though it is not possible to measure the spins directly at the current stage, we show that several quantities are able to be probed through the decay distributions with cascade decays. 
In particular, $B_s^0 \to \Lambda(\to p \pi^-) \overline{\Lambda}(\to \overline{p} \pi^+)$ provide excellent opportunities  as its final state particles are all charged.

We have estimated these quantities through the $^3P_0$ model and chiral selection rule within the SM. The decaying branching ratios are found to be 
${\cal B}(B^0 \to p \overline{p})=1.27 \times 10 ^{-8} $ and ${\cal B}(B^+ \to p \overline{\Lambda})=2.31 \times 10 ^{-7} $, which are consistent with the current experimental data. On the other hand, ${\cal B}(B_s^0 \to \Lambda \overline{\Lambda})$ is estimated to be around $7.4\times 10 ^{-7}$, which is about 60 times larger than ${\cal B}(B^0 \to p \overline{p})$, making it promising to be observed in the near future.  
In addition, we show that $\langle A_{dir}\rangle$ are tiny for $B_s^0$ due to the violent oscillation between $B_s^0$ and $\overline{B}_s^0$. 

We suggest the future experiments to visit $B_s^0 \to 
\Lambda(\to p \pi^-)
\overline{\Lambda} (\to \overline{ p}  \pi^+) $, where the time-averaged CP-odd observables are estimated to be $(\langle \alpha_w \rangle ,\langle\beta_w\rangle )= (-2.2\%, 0.5\%) $ or
$(\langle \alpha_w \rangle ,\langle\beta_w\rangle )= (3.1\%, -0.7\%) $, depending on the sign of the strong phases.

\section*{Acknowledgments}
This work is supported in part by the National Key Research and Development Program of China under Grant No. 2020YFC2201501 and  the National Natural Science Foundation of China (NSFC) under Grant No. 12147103 and 12205063.


\begin{thebibliography}{99}


\bibitem{CPNP}
M.~Saur and F.~S.~Yu,
Sci. Bull. \textbf{65}, 1428 (2020);
C.~Q.~Geng and C.~W.~Liu,
JHEP \textbf{11}, 104 (2021);
X.~G.~He, J.~Tandean and G.~Valencia,
Sci. Bull. \textbf{67}, 1840 (2022);
X.~G.~He and J.~P.~Ma,
Phys. Lett. B \textbf{839}, 137834 (2023);
Z.~H.~Zhang and J.~J.~Qi,
Eur. Phys. J. C \textbf{83},  133 (2023);
Z.~H.~Zhang,
Phys. Rev. D \textbf{107},  L011301 (2023).



\bibitem{LHCb:2017hkl}
R.~Aaij \textit{et al.} [LHCb],
JHEP \textbf{03}, 059 (2018); 
R.~Aaij \textit{et al.} [LHCb],
JHEP \textbf{06}, 084 (2018).


\bibitem{Gronau:1990ra}
M.~Gronau and D.~London,
Phys. Lett. B \textbf{253}, 483 (1991);
M.~Gronau and D.~Wyler,
Phys. Lett. B \textbf{265}, 172 (1991);
D.~Atwood, I.~Dunietz and A.~Soni,
Phys. Rev. Lett. \textbf{78}, 3257 (1997);
D.~Atwood, I.~Dunietz and A.~Soni,
Phys. Rev. D \textbf{63}, 036005 (2001);
Y.~Grossman, Z.~Ligeti and A.~Soffer,
Phys. Rev. D \textbf{67}, 071301 (2003); 
A.~Giri, Y.~Grossman, A.~Soffer and J.~Zupan,
Phys. Rev. D \textbf{68}, 054018 (2003);
A.~Bondar and A.~Poluektov,
Eur. Phys. J. C \textbf{47}, 347 (2006).

\bibitem{CKM} 
CKMfitter group, J. Charles {\it et al.}, 
Phys. Rev. D {\bf 91} 073007 (2015);
UTfit collaboration, M. Bona {\it et al.}, 
JHEP {\bf 10},  081 (2006).

\bibitem{Geng:2022osc}
A.~K.~Giri, R.~Mohanta and M.~P.~Khanna,
Phys. Rev. D \textbf{65}, 073029 (2002);
S.~Zhang, Y.~Jiang, Z.~Chen and W.~Qian,
arXiv:2112.12954 [hep-ph];
C.~Q.~Geng, X.~N.~Jin, C.~W.~Liu, Z.~Y.~Wei and J.~Zhang,
Phys. Lett. B \textbf{834}, 137429 (2022);
S.~Roy, N.~G.~Deshpande, A.~Kundu and R.~Sinha,
arXiv:2303.02591 [hep-ph].



\bibitem{BCP}
B. Aubert et al. [BaBar], Phys. Rev. Lett. {\bf 86}, 2515 (2001);
K. Abe et al. [Belle], Phys. Rev. Lett. {\bf 87}, 091802 (2001).

\bibitem{pdg}
R.~L.~Workman \textit{et al.} [Particle Data Group],
PTEP \textbf{2022}, 083C01 (2022).


\bibitem{LHCb:2014kcb}
R.~Aaij \textit{et al.} [LHCb],
JHEP \textbf{07}, 041 (2014);
R.~Aaij \textit{et al.} [LHCb],
Phys. Rev. Lett. \textbf{116},  191601 (2016);
R.~Aaij \textit{et al.} [LHCb],
Phys. Rev. Lett. \textbf{122},  211803 (2019).

\bibitem{DmesonCP}
Z.~z.~Xing,
Mod. Phys. Lett. A \textbf{34},  1950238 (2019);
Y.~Grossman and S.~Schacht,
JHEP \textbf{07}, 020 (2019);
H.~Y.~Cheng and C.~W.~Chiang,
Phys. Rev. D \textbf{100},  093002 (2019);
M.~Chala, A.~Lenz, A.~V.~Rusov and J.~Scholtz,
JHEP \textbf{07}, 161 (2019);
A.~Dery and Y.~Nir,
JHEP \textbf{12}, 104 (2019);
L.~Calibbi, T.~Li, Y.~Li and B.~Zhu,
JHEP \textbf{10}, 070 (2020);
H.~Y.~Cheng and C.~W.~Chiang,
Phys. Rev. D \textbf{104}, 073003 (2021).



\bibitem{LHCb:2016rja}
R.~Aaij \textit{et al.} [LHCb],
JHEP \textbf{05}, 081 (2016).


\bibitem{PQCD}
C.~Q.~Geng, C.~W.~Liu and T.~H.~Tsai,
Phys. Lett. B \textbf{815}, 136125 (2021);
J.~J.~Han, Y.~Li, H.~n.~Li, Y.~L.~Shen, Z.~J.~Xiao and F.~S.~Yu,
Eur. Phys. J. C \textbf{82},  686 (2022);
C.~W.~Liu and C.~Q.~Geng,
JHEP \textbf{01}, 128 (2022).

\bibitem{private}
Private communication with Jia-Jie Han and Fu-Sheng Yu; to be appear on arXiv.


\bibitem{ProductionRate}
R.~Aaij \textit{et al.} [LHCb],
Phys. Rev. D \textbf{85}, 032008 (2012);
R.~Aaij \textit{et al.} [LHCb],
Phys. Rev. D \textbf{99},  052006 (2019);
R.~Aaij \textit{et al.} [LHCb],
Phys. Rev. D \textbf{100},  031102 (2019).


\bibitem{Chua:2022wmr}
C.~K.~Chua,
Phys. Rev. D \textbf{95},  096004 (2017);
C.~K.~Chua,
Phys. Rev. D \textbf{106}, 036015 (2022).



\bibitem{LHCb:2016nbc}
R.~Aaij \textit{et al.} [LHCb],
JHEP \textbf{04}, 162 (2017).



\bibitem{LHCb:2022lff}
 [LHCb],
arXiv:2206.06673 [hep-ex].

\bibitem{Jin:2021onb}
X.~N.~Jin, C.~W.~Liu and C.~Q.~Geng,
Phys. Rev. D \textbf{105},  053005 (2022).



\bibitem{Cheng:2001ub}
H.~Y.~Cheng and K.~C.~Yang,
Phys. Rev. D \textbf{65}, 054028 (2002)
[erratum: Phys. Rev. D \textbf{65}, 099901 (2002)].


\bibitem{Chernyak:1990ag}
V.~L.~Chernyak and I.~R.~Zhitnitsky,
Nucl. Phys. B \textbf{345}, 137 (1990).


\bibitem{Cheng:2014qxa}
H.~Y.~Cheng and C.~K.~Chua,
Phys. Rev. D \textbf{91},  036003 (2015).

\bibitem{BVV}
C.~Q.~Geng and C.~W.~Liu,
Phys. Lett. B \textbf{825}, 136883 (2022).



\bibitem{Micu:1968mk}
L.~Micu,
Nucl. Phys. B \textbf{10}, (1969).

\bibitem{LeYaouanc:1972vsx}
A.~Le Yaouanc, L.~Oliver, O.~Pene and J.~C.~Raynal,
Phys. Rev. D \textbf{8}, (1973).

\bibitem{Ackleh:1996yt}
E.~S.~Ackleh, T.~Barnes and E.~S.~Swanson,
Phys. Rev. D \textbf{54}, (1996).


\bibitem{Segovia:2012cd}
J.~Segovia, D.~R.~Entem and F.~Fern\'andez,
Phys. Lett. B \textbf{715}, 322 (2012).





\bibitem{Simonov:2011cm}
Y.~A.~Simonov,
Phys. Rev. D \textbf{84}, 065013 (2011).


\bibitem{BESIII:2019hdp}
M.~Ablikim \textit{et al.} [BESIII],
Phys. Rev. Lett. \textbf{124}, 042001 (2020).


\bibitem{Jpsi}
C.~Q.~Geng, C.~W.~Liu and J.~Zhang,
arXiv:2306.02138 [hep-ph].



\bibitem{Liu:2022pdk}
C.~W.~Liu and C.~Q.~Geng,
Chin. Phys. C {\bf 47}, 014104 (2023).




\end{thebibliography}
\end{document}